\documentclass[10pt]{article}
\title{\bf A Set of Identities for a Class of Alternating Binomial Sums Arising in Computing Applications} 
\author{Mark W. Coffey\\
                     \\
Department of Physics\\
Colorado School of Mines\\
Golden, CO  80401  USA\\
\{mcoffey@mines.edu\} } 
\date{}
\begin{document}
\maketitle
\baselineskip=15 pt
\begin{abstract}

\noindent
We perform certain alternating binomial summations with parameters that
occur in 
the analysis of algorithms.  A combination of integral and special function 
and special number representations is used.  The results are sufficiently
general to subsume several previously known cases.  Extensions of the method
are apparent and are outlined.

\end{abstract}
 
\vspace{2.0cm}
\baselineskip=15pt
\centerline{\bf Key words and phrases}
\medskip 

\noindent
binomial summation, Stirling numbers, Beta function, polygamma function, 
generalized harmonic numbers, generating function, Pochhammer symbol, binomial coefficient, Bell polynomial 

 
\baselineskip=15pt
\pagebreak
\medskip
\centerline{\bf \Large Introduction}
\medskip

Alternating binomial sums arise frequently in computer science and data 
processing in the design and analysis of algorithms (e.g., 
\cite{flajolet,kirschprod,kirsch94,knuth,prod}).
The asymptotic form of such sums is often of interest in connection with
determining the average- or worst-case run time.  Because of the sign
alternation of the summands, there may be substantial cancellation, masking
the dominant behaviour.  Recently there has been additional interest in
certain alternating binomial sums and their connection with harmonic numbers
and representation in terms of the Bell polynomials \cite{kirsch,larcombe05}.
In fact, the question of a broader range of validity of such identities has
been posed \cite{larcombe05}.  In this article, we demonstrate a method
to reach alternating binomial sum representations with a domain extended
to the complex plane.  

Our approach is to obtain exact analytic relations.  To these, known
asymptotic relations may be applied if desired.  Our results help to 
elucidate the connections between certain Bell polynomial representations,
generalized harmonic numbers, Stirling numbers, and special values of the
polygamma functions, and should be helpful in the analysis of either
deterministic or probabilistic algorithms.  Various integral 
representations provide a convenient centerpoint of our development, but
this is by no means necessary.  There are many complementary approaches,
including the use of finite difference operators \cite{larcombe05,comtet}.
Afterall, $\Delta^n f(x) = (E-I)^n f(x) = (-1)^n \sum_{k=0}^n (-1)^k 
{n \choose k} f(x+k)$ where $\Delta f(x) \equiv f(x+1) -f(x)$, 
$Ef(x) \equiv f(x+1)$, and $I$ is the identity operator.
A way to think of the underlying combinatorics is in terms of number
partitioning needed in the course of differentiating composite functions
\cite{wcyang}.  

In this paper, we calculate the alternating binomial sums
$$S(x,N,m) \equiv \sum_{k=0}^N {N \choose k} {{(-1)^k} \over {(x+k)^m}},
\eqno(1)$$
in multiple fashion for positive integers $N$ and $m$ and complex 
$x \in C/\{0,-1,$ $-2,\ldots,-N\}$.  The special cases $x=\pm K$ for $K$ a 
positive integer recover results of Kirschenhofer \cite{kirsch} and
Larcombe et al. \cite{larcombe05}.   The special case of $x=1$ in Eq. (1)
has further applications in quantum information science \cite{coffeyjpa06}
and this case is evident in our scheme.  We point out additional special
cases for $x$ a rational number. 
 
Our work is more illustrative than exhaustive as there are many possible
extensions.  Especially when the summand in question contains a function
for which Lemma 1 below applies, there will always be an equivalent
representation in terms of the Bell polynomials $Y_m(x_1,\ldots,x_m)$. 
Moreover, as Eq. (10) below demonstrates, any time an expression contains
Stirling numbers of the first kind, these may be replaced with Bell 
polynomials with generalized harmonic number arguments.
Of the Bell polynomials, we note that they may be written as a lower
triangular determinant save for a superdiagonal of $-1$'s:
$$Y_n(x_1,\ldots,x_n)=\left|\begin{array}{cccccc}
x_1 &  -1  &   0 &  0  & \ldots &  0 \\
x_2 & x_1  &  -1 &  0  & \ldots &  0 \\
x_3 & 2x_2 &  x_1 &-1  & \ldots &  0 \\
x_4 & 3x_3 &  3x_2 & x_1 & \ldots & 0 \\
\vdots & \vdots & \vdots & \vdots & \ldots & 0\\
\vdots & \vdots & \vdots & \vdots & \ldots & -1\\
x_n & {{n-1} \choose 1}x_{n-1} & {{n-1} \choose 2}x_{n-2} &{{n-1} \choose 3}
x_{n-3} & \ldots & {{n-1} \choose {n-1}} x_1\end{array} \right|.  \eqno(2)$$
They satisfy the recursion relation
$$Y_{n+1}(x_1,x_2,\ldots,x_{n+1})=\sum_{k=0}^n {n \choose k} Y_{n-k}(x_1,x_2,
\ldots,x_{n-k})x_{k+1}, \eqno(3a)$$
and
$$Y_n(x_1+y_1,\ldots,x_n+y_n)=\sum_{k=0}^n {n \choose k}Y_{n-k}(x_1,
x_2,\dots,x_{n-k})Y_k(y_1,y_2,\ldots,y_k),$$
$$~~~~~~~~~~~~~~~~~~~~~~~~~~~~~~~~~~~~~~~~~~~~~~~~~~~~~~~~~~~~~~~~~~~~ n \geq 0.  
\eqno(3b)$$
Reference \cite{schimming} contains a background section on the Bell
polynomials, wherein Proposition 2 repeats the well known determinant 
expression (2).  Those authors also denote by $B_n^-$ ``inverse Bell polynomials"
that have usually been called logarithmic polynomials $L_n$ \cite{comtet}
(p. 140).  Proposition 1 of Ref. \cite{schimming} covers the parity of Bell
polynomials when the even- or odd-indexed variables are put to zero and may
instead by obtained from the determinantal expression (2).
For further information on $Y_n$ we refer to standard works \cite{comtet,riordan1,riordan2}.

Analytic number theory is an additional
field where significant alternating sums occur and these applications should
not be overlooked.  In fact, we take up an important such instance elsewhere.

%
%

\bigskip
\centerline{\bf \Large Summary of results and preparation}
\medskip

Let $N>0$ and $m>0$ be integers, $x \in C/\{0,-1,-2,\ldots,-N\}$, $\Gamma$
the Gamma function, $(a)_n=\Gamma(a+n)/\Gamma(a)$ the Pochhammer symbol,
$B$ the Beta function, and $_pF_q$ the generalized hypergeometric function
\cite{andrews,krupnikov}.  Let $s(n,k)$ and $S(n,k)$
be Stirling numbers of the first and second kind, respectively \cite{nbs,comtet,gkp,riordan1,riordan2}.  We have
\newline{\bf Proposition 1}
$$S(x,N,m) = {1 \over x^m} ~_{m+1}F_m(x,\ldots,x,-N;x+1,\ldots,x+1;1) \eqno(4)$$
$$= {{(-1)^{m-1}} \over {(m-1)!}} \left({\partial \over {\partial x}}
\right)^{m-1} {{N! \Gamma(x)} \over {\Gamma(N+x+1)}} \eqno(5)$$
$$={1 \over {(m-1)!}}\int_0^\infty t^{m-1} e^{-xt}(1-e^{-t})^N dt \eqno(6)$$
$$={{2^{N+m}} \over {(m-1)!}}\int_0^\infty w^{m-1} e^{-(2x+N)w} \sinh^N w ~dw
\eqno(7)$$
$$={{N!} \over {(m-1)!}}\sum_{n=N}^\infty {{S(n,N)} \over {n!}} {{(n+m-1)!}
\over {(x+N)^{n+m}}} \eqno(8)$$
$$=\sum_{n=m-1}^\infty {{(-1)^n} \over {n!}} s(n,m-1)B(N+n+1,x) \eqno(9)$$
$$={{(-1)^{m-1}} \over {(m-2)!}} \sum_{n=m-1}^\infty {{B(n+N+1,x)} \over n}
Y_{m-2}[H_{n-1},-H_{n-1}^{(2)},2!H_{n-1}^{(3)},\ldots,$$
$$~~~~~~~~~~~~~~~~~~~~~~~~~~~~~~~~~~~~~~~~~(-1)^{m-1}(m-3)!H_{n-1}^{(m-2)}], 
\eqno(10)$$
$$={{(-1)^{m-1}} \over {(m-1)!}} {{N!} \over {(x)_{N+1}}} Y_{m-1}\left[
g(x),g'(x),\ldots,g^{(m-1)}(x)\right].  \eqno(11)$$
In Eq. (11), 
$$g(x) = \psi(x)-\psi(x+N+1)=-\sum_{k=0}^N {1 \over {x+k}}, \eqno(12a)$$
$$g^{(\ell)}(x)=\psi^{(\ell)}(x)-\psi^{(\ell)}(x+N+1)=-(-1)^\ell \ell !
\sum_{k=0}^N {1 \over {(x+k)^{\ell+1}}}, \eqno(12b)$$
where $\psi=\Gamma'/\Gamma$ is the digamma function and $\psi^{(j)}$ the
polygamma function \cite{nbs}.  The summation expressions in Eq. (12) follow
by the use of the functional equations of these functions.  In Eq. (10),
$H_n^{(\ell)}$ are generalized harmonic numbers, 
$$H_n^{(r)}=\sum_{k=1}^n {1 \over k^r}={{(-1)^{r-1}} \over {(r-1)!}}
\left[\psi^{(r-1)}(n+1)-\psi^{(r-1)}(1)\right].  \eqno(13)$$

The proof of the equivalence given in Eq. (11) makes use of the following.
{\newline \bf Lemma 1}.  For differentiable functions $f$ and $g$ such that
$f'(x)=f(x)g(x)$, assuming all higher order derivatives exist, we have
$$\left({d \over {dx}}\right)^j f(x) = f(x) Y_j\left[g(x),g'(x),\ldots,g^{(j-1)}
(x)\right].  \eqno(14)$$

Proof of Lemma 1.  Under the premise, $g(x)=(d/dx) \ln f(x)$ whenever
$f \neq 0$ and $(d/dx)^j f=f\exp(-\ln f) (d/dx)^j \exp(\ln f)$.  The
conclusion then follows as a special case of the Fa\`{a} di Bruno formula
for the derivative of a composite function.  The result
extends to $x \in C$ when a branch cut for the logarithm is taken from the
origin to the point at infinity.

{\bf Remarks and an example}.  (i) The (exponential) complete Bell polynomials
$Y_n$ may be obtained as a sum over the (exponential) partial Bell polynomials
$B_{n,k}$:  $Y_n(x_1,\ldots,x_n)=\sum_{k=1}^n B_{n,k}(x_1,x_2,\ldots,x_{n-k+1})$,
$Y_0=1$, $Y_1(x_1)=x_1$.
(ii) The condition of the Lemma occurs often in practice and a nice example is
given by the Gamma function with $\Gamma'=\Gamma \psi$ (\cite{comtet}, p. 175).
Then $\Gamma^{(n)}(1)=Y_n(-\gamma,x_2,x_3,\ldots,x_n)=\int_0^\infty e^{-x}
\ln^n x ~dx$, where $\gamma=-\psi(1)$ and $x_j=(-1)^j (j-1)!\zeta(j)$ with
$\zeta(s)$ the Riemann zeta function.

(iii) The equality of Eqs. (3) and (4) is obvious.  The original binomial
series form (1) may be returned from Eq. (4) by the following two steps.
First, depending upon whether $N$ is even or odd we substitute into Eq. (4)
either the expansion (\cite{grad}, p. 25)
$$\sinh^{2n} x={{(-1)^n} \over 2^{2n}}\left[\sum_{k=0}^{n-1}(-1)^{n-k} 2
{{2n} \choose k} \cosh 2(n-k)x + {{2n} \choose n}\right], \eqno(15a)$$
or 
$$\sinh^{2n-1} x ={{(-1)^{n-1}} \over 2^{2n-2}}\sum_{k=0}^{n-1} (-1)^{n+k-1} 2 {{2n-1} \choose k} \sinh (2n-2k-1)x. \eqno(15b)$$
We then apply tabulated integrals (\cite{grad}, p. 360) to find Eq. (1).

(iv) The upper limit on the summation in Eq. (1) could just as well be put
to $\infty$ due to the property ${n \choose k} = 0$ for $k > n$.

(v) We are using notation for the Stirling numbers as followed by Comtet
\cite{comtet} and Riordan \cite{riordan1} and the reader should be aware of
other conventions.  
Indeed the notation for these numbers has never been standardized \cite{nbs}.
(vi) The Stirling numbers of the second kind are of rank one
while of the first kind are of rank two.  I.e., the latter numbers require
a double summation in order to be expressed in terms of elementary factors
\cite{comtet}.
(vii) We do not require them here, but mention that asymptotic forms of the
Stirling numbers and functions are known. 

\medskip
\centerline{\bf \Large Proof of Proposition 1}
\medskip

We now proceed systematically through the list of equivalences given in
Proposition 1.  In writing Eq. (1) in the form (4) we use the power series
form of $_{m+1}F_m$ \cite{andrews} and apply $(x)_k/(x+1)_k=x/(x+k)$.
The terminating hypergeometric series in Eq. (4) is $m+N$-balanced since the
sums of numerator and denominator parameters differ by this positive integer.

We next recognize that
$$\sum_{k=0}^N {N \choose k}{{(-1)^k} \over {x+k}}={{N! \Gamma(x)} \over
{\Gamma(x+N+1)}}={{N!} \over {(x)_{N+1}}} = {{N!} \over {x(x+1)_N}}=B(x,N+1),
\eqno(16)$$
that is equivalent to a partial fractional decomposition.  Then
$$\left({\partial \over {\partial x}} \right)^{m-1} {{N! \Gamma(x)} \over {\Gamma(N+x+1)}} = (-1)^{m-1}(m-1)!S(x,N,m)$$
$$=\left({\partial \over {\partial x}}\right)^{m-1} \sum_{k=0}^N  {N \choose k}
(-1)^k \int_0^\infty e^{-(x+k)t} ~dt$$
$$=(-1)^{m-1} \sum_{k=0}^N {N \choose k} (-1)^k \int_0^\infty t^{m-1}
e^{-(x+k)t} ~dt$$
$$=(-1)^{m-1}\int_0^\infty t^{m-1} e^{-xt} (1-e^{-t})^N dt. \eqno(17)$$
In the above the interchange of differentiation and integration is justified
by the absolute convergence of the integral.  We have shown the equality of
Eq. (1) and (4)-(6).  Equation (7) follows from (6) by using the definition 
of the hyperbolic sine function in terms of exponentials.  

To obtain the form of Eq. (8) we write Eq. (6) as
$$S(x,N,m)={1 \over {(m-1)!}}\int_0^\infty t^{m-1} e^{-(x+N)t} (e^t-1)^N dt,
\eqno(18)$$
and apply a generating function for the Stirling numbers of the second kind
\cite{nbs}:
$$(e^x-1)^m=m!\sum_{n=m}^\infty S(n,m) {x^n \over {n!}}.  \eqno(19)$$
In obtaining Eq. (9) we first make the change of variable $v(t)=1-e^{-t}$ in
Eq. (6), giving
$$S(x,N,m)={{(-1)^{m-1}} \over {(m-1)!}}\int_0^1 v^N (1-v)^{x-1} \ln^{m-1}(1-v)
~dv.  \eqno(20)$$
We then use a generating function for Stirling numbers of the first kind,
$$\ln^m(1+x)=m!\sum_{n=m}^\infty s(n,m) {x^n \over {n!}}, \eqno(21)$$
and carry out the integration with the Beta function.  

For Eq. (10) we apply Theorem B of Ch. V of Ref. \cite{comtet} (p. 217) for
the unsigned Stirling number of the first kind.  We write this result in the
form
$$|s(n+1,k+1)|=(-1)^{n+k}s(n+1,k+1)={{n!} \over {k!}}Y_k[H_n,-H_n^{(2)},2!
H_n^{(3)},\ldots,$$
$$~~~~~~~~~~~~~~~~~~~~~~~~~~~~~~~~~~~~~~~~~~~~~~~~~~~~~~(-1)^{k-1}(k-1)!H_n^{(k)}].  
\eqno(22)$$
We then substitute for $s(n,m-1)$ in Eq. (9).

In order to obtain the form (11) we use Eq. (5) and apply Lemma 1 with the
function $f(x)=N!/(x)_{N+1}$, such that
$${d \over {dx}} f(x) = f(x) [\psi(x)-\psi(x+N+1)],  \eqno(23)$$
providing the function $g(x)$ presented in Eq. (12a).  Hence with this particular
$g(x)$ we have 
$$\left({d \over {dx}}\right)^j f(x) = f(x) Y_j\left[g(x),g'(x),\ldots,g^{(j-1)}
(x)\right],  \eqno(24)$$
and the rest of Proposition 1 follows.

{\bf Remarks}.  (i)  There are many other variations on the possible generating
functions that may be introduced into the integrand of Eq. (6) to produce
equivalent forms with the Stirling numbers.  Ref. \cite{coffeyjpa06}
provides examples of these alternatives. (ii)  The lower limit of summation in
Eqs. (8) and (9) could just as well be put to $0$ due to the property 
$s(n,k)=S(n,k)=0$ for $n < k$.  (iii)  Equations (9) and (20) 
make it strikingly apparent how the case of $x=1$ is special, when the Beta
function is no longer required.  (iv)  Single or repeated integration by
parts in Eq. (6) permits the derivation of other forms of $S(x,N,m)$ and
of recursion relations for these sums.  For instance, from $e^{-xt}=-(1/x)
(d/dt)e^{-xt}$ and $[e^t/(N+1)](d/dt)(1-e^{-t})^{N+1}$ we immediately have
$$S(x,N,m)={1 \over x}[S(x,N,m-1)+NS(x-1,N-1,m)], \eqno(25a)$$
and
$$S(x,N,m)={1 \over {N+1}}[(x-1)S(x-1,N+1,m)-S(x-1,N+1,m-1)],  \eqno(25b)$$
respectively.  These relations have been obtained subject to Re $x >0$ for
Eq. (25a) and Re $x >1$ for Eq. (25b).

\medskip
\centerline{\bf \Large Special cases}
\medskip

We very briefly mention cases where $x$ is an integer or a rational number 
in Eqs. (1) and (11). When $x=1$ we have for Eq. (12)
$$g(1)=\psi(1)-\psi(N+2)=-\sum_{k=1}^{N+1} {1 \over k} \equiv -H_{N+1}, 
\eqno(26a)$$
and
$$g^{(\ell)}(1)=-(-1)^\ell \ell !\sum_{k=1}^{N+1} {1 \over {k^{\ell+1}}}
\equiv -(-1)^\ell \ell ! H_{N+1}^{(\ell+1)}, \eqno(26b)$$
where $H_p^{(r)}$ are generalized harmonic numbers.  These polygamma values
are well known to relate to differences of the Riemann zeta function at
integer argument, as $\psi^{(\ell)}(1)=-(-1)^\ell \ell ! \zeta(\ell+1)$.
Here $f(1)=N!/(N+1)!=1/(N+1)$ and then
$$\left. \left({d \over {dx}}\right)^j f(x)\right|_{x=1} = {1 \over {N+1}}
Y_j[-H_{N+1},H_{N+1}^{(2)},-2!H_{N+1}^{(3)},\ldots, (-1)^j(j-1)!H_{N+1}^{(j)}].
\eqno(27)$$

When $x$ is a rational number it is possible to re-express the necessary
derivatives $g^{(\ell)}(x)$.  This can be done either in terms of the
polygamma function or in terms of the Hurwitz and Riemann zeta functions
(e.g., \cite{kolbig}).  For instance we have $\psi^{(\ell)}(1/2)=(-1)^{\ell+1}
\ell ! (2^{\ell+1}-1)\zeta(\ell+1)$.  

When $x$ is an integer we are able to express the sums of Eq. (12) in terms 
of generalized harmonic numbers.  For instance we have for $K$ a positive integer
$$g^{(\ell)}(-K)=(-1)^{\ell+1}\ell !\sum_{\stackrel{k=0}{k \neq K}}^N {1 \over
{(k-K)^{\ell+1}}}=(-1)^{\ell+1}\ell ![H_{N-K}^{(\ell+1)} + (-1)^\ell 
H_K^{(\ell+1)}], \eqno(28a)$$
and 
$$g^{(\ell)}(K)=(-1)^{\ell+1} \ell ![H_{N+K}^{(\ell+1)} - H_{K-1}^{(\ell+1)}]. \eqno(28b)$$


\medskip
\centerline{\bf \Large Extensions}
\medskip

The approach of this article may be extended to a great many other integrals.
We outline some of this using the Beta function as the base.  However,
one could just as well apply the techniques to the confluent hypergeometric
function $_1F_1$, the Gauss hypergeometric function $_2F_1$, and then to $_pF_q$
more generally.

We consider
$$B(x,y)=\int_0^1 u^{x-1}(1-u)^{y-1}du = 2\int_0^{\pi/2} \sin^{2x-1} \phi
\cos^{2y-1} \phi ~d\phi = B(y,x),$$
$$~~~~~~~~~~~~~~~~~~~~~~~~~~~~~~~~~~~~~~~~~~~~~~~~~~~~~~~~~~~~~~ \mbox{min}[\mbox{Re} ~x, \mbox{Re} ~y]>0, \eqno(29)$$
so that
$$S(x,y,m,n) \equiv \int_0^1 u^{x-1} (1-u)^{y-1}\ln^{m-1}u \ln^{n-1}(1-u)
~du $$
$$=\left({\partial \over {\partial x}}\right)^{m-1}
\left({\partial \over {\partial y}}\right)^{n-1} B(x,y).  \eqno(30)$$
Just as generalized binomial expansion gives
$$B(x,y)=\sum_{j=0}^\infty (-1)^j {{y-1} \choose j} {1 \over {x+j}},
\eqno(31)$$
we have
$$S(x,y,m,n)=(-1)^{m-1} (m-1)!\sum_{j=0}^\infty {1 \over {j!}}\left( 
{\partial \over {\partial y}}\right)^{n-1} (1-y)_j {1 \over {(x+j)^m}}.
\eqno(32)$$
In obtaining this equation we wrote ${{y-1} \choose j}=(-1)^j(1-y)_j/j!$.
Of course the sum terminates when $y-1$ is a positive integer and we
again exclude nonpositive integer values for $x$.
The Pochhammer polynomial occurring in Eq. (32) has derivative
$${d \over {dy}}(1-y)_j=(1-y)_j[\psi(1-y)-\psi(j+1-y)].  \eqno(33)$$
Therefore, for instance, Lemma 1 may be used in finding all higher order
derivatives.  

We may next change variable in Eq. (30) to obtain
$$S(x,y,m,n) = (-1)^{n-1} \int_0^\infty v^{n-1} \ln^{m-1} (1-e^{-v})
(1-e^{-v})^{x-1} e^{-yv} dv. \eqno(34)$$ 
Otherwise, we may first put $z=1/u$ in Eq. (29) and $v=\ln z$, giving
$$B(x,y)=\int_0^\infty e^{-(x+y-1)v} (e^v-1)^{y-1} dv.  \eqno(35)$$
Then we have
$$S(x,y,m,n)=\int_0^\infty e^{-(x+y-1)v}(e^v-1)^{y-1}\left[(-1)^{m-1} v^{m-1}
\ln^{n-1} (e^v-1)\right.$$
$$\left.~~~~~~~~~~~~~~~~~~~~~~~~~~~~~~~~~~~~~~~~~~~~~~~ + (-1)^{m+n} v^{m+n-2}
\right ]dv.  \eqno(36)$$
Equations (34) and (36) are forms suitable for re-expression as a summation
with Stirling number coefficients.

By the same token, one may repeatedly integrate the Beta function, thereby
obtaining binomial summation expressions for integrals of the form
$$I(x,y,p,m)=\int_0^1 {{(1-u^p)^x u^{y-1}} \over {\ln^m u}}du, \eqno(37)$$
for Re $x>0$, Re $y>0$, and $m$ an integer.  In this way, we obtain extensions
of tabulated integrals such as given in Sections 4.267 and 4.268 of Ref.
\cite{grad}.

{\bf Final remarks}.  We could also develop series representations with
Stirling number coefficients using divided difference formulas.  For example,
we have \cite{nbs}
$$\left({d \over {dx}}\right)^m f(x)=m!\sum_{n=m}^\infty {{s(n,m)} \over {n!}}
\Delta^n f(x), \eqno(38a)$$
and
$$\Delta^m f(x) = m! \sum_{n=m}^\infty {{S(n,m)} \over {n!}} f^{(n)}(x),
\eqno(38b)$$
these formulas also exhibiting the inverse relations possible with the
Stirling numbers.  Especially for the Gamma and hence the Beta function,
the finite differences and derivatives are relatively easily determined,
due to their respective functional equations.

\bigskip
\centerline{\bf \Large Acknowledgement}   
\medskip
This work was partially supported by Air Force contract number FA8750-04-1-0298.



\pagebreak


\begin{thebibliography}{99}
\bibitem{nbs}M. Abramowitz and I. A. Stegun,
{Handbook of Mathematical Functions, Washington, National Bureau of Standards
(1964).}
\bibitem{andrews}G. E. Andrews, R. Askey, and R. Roy,
{Special functions, Cambridge University Press (1999)}.
\bibitem{coffeyjpa06}M. W. Coffey,
{One integral in three ways:  moments of a quantum distribution,
J. Phys. A {\bf 39}, 1425-1431 (2006).}
\bibitem{comtet}L. Comtet,
{Advanced Combinatorics, D. Reidel (1974).}
\bibitem{grad}I. S. Gradshteyn and I. M. Ryzhik,
{Table of Integrals, Series, and Products, Academic Press, New York (1980).}
\bibitem{flajolet}P. Flajolet and R. Sedgewick,
{Digital search trees revisited, SIAM J. Comput. {\bf 15}, 748-767 (1986).}
\bibitem{gkp}R. L. Graham, D. E. Knuth, and O. Patashnik,
{Concrete Mathematics, 2nd ed., Addison Wesley (1994).}
\bibitem{kirsch}P. Kirschenhofer,
{A note on alternating sums, El. J. Comb. {\bf 3}, \#R7 (1996).}
\bibitem{kirschprod}P. Kirschenhofer and H. Prodinger,
{Approximate counting:  an alternative approach, Inf. Th. Appl. {\bf 25},
43-48 (1991).}
\bibitem{kirsch94}P. Kirschenhofer and H. Prodinger,
{The path length of random skip lists, Acta Inf. {\bf 31}, 775-792 (1994).}
\bibitem{knuth}D. E. Knuth,
{The Art of Computer Programming, Vol. 3, Addison Wesley (1973).}  
\bibitem{kolbig}K. S. K\"{o}lbig,
{The polygamma function $\psi^{(k)}(x)$ for $x=1/4$ and $x=3/4$,
J. Comput. Appl. Math. {\bf 75}, 43-46 (1996).}
\bibitem{krupnikov}E. D. Krupnikov and K. S. K\"{o}lbig,
{Some special cases of the generalized hypergeometric function $_{q+1}F_q$,
J. Comput. Appl. Math. {\bf 78}, 79-95 (1997).}
\bibitem{larcombe05}P. J. Larcombe, M. E. Larsen, and E. J. Fennessey,
{On two classes of identities involving harmonic numbers, Util. Math. 
{\bf 67}, 65-80 (2005).}
\bibitem{prod}H. Prodinger,
{Combinatorics of geometrically distributed random variables:  Left-to-right
maxima, Discrete Math. {\bf 153}, 253-270 (1996).}
\bibitem{riordan1}J. Riordan,
{An introduction to combinatorial analysis, Wiley (1958).}
\bibitem{riordan2}J. Riordan,
{Combinatorial identities, Wiley (1968).}
\bibitem{schimming}R. Schimming and W. Strampp,
{Differential polynomial expressions related to the Kadomtsev-Petviashvili and
Korteweg-de Vries hierarchies, J. Math. Phys. {\bf 40}, 2429-2444 (1999).}
\bibitem{wcyang}W. C. Yang,
{Derivatives are essentially integer partitions, Discrete Math. {\bf 222},
235-245 (2000).}
\end{thebibliography}
\end{document}